# Resonant pump intensity dependence of luminescence achieved via Floquet engineering

I. S. Pashkevich[1], I. V. Doronin[1,2,3,*], A. A. Zyablovsky[1,2,4], E. S. Andrianov[1,2,4]

[1]Moscow Institute of Physics and Technology, 9 Institutskiy pereulok, Moscow, Russia, 141700

[2]Dukhov Research Institute of Automatics, 22 Sushchevskaya, Moscow, Russia, 127055

[3]Institute of Spectroscopy of Russian Academy of Sciences, Fizicheskaya str. 5, Troitsk, Moscow, Russia, 108840

[4]Institute for Theoretical and Applied Electromagnetics, 13 Izhorskaya, Moscow, Russia, 125412

[*]ildoron2@gmail.com

**Abstract**

We discover an unexpected behavior in a hybrid system composed of cavity strongly coupled to molecules and subjected to high intensity coherent pumping. We show that if the frequency of the pumping wave is close to polariton transitions in the hybrid system, non-monotone dependence of fluorescence and scattering amplitudes on pump intensity with a narrow resonance-like response occurs. We demonstrate that this phenomenon occurs due to hybridization of lower and upper polaritons with substantially different excitation numbers caused by the pumping affecting polariton states of the system. This occurs when the Rabi interaction with coherent field is comparable to the field-matter coupling constant which in turn needs to be sufficient for the manifestation of strong coupling. This non-monotonic dependence of the fluorescence and scattering amplitudes on pump rate intensity pave the way for creation of nonlinear optical devices.

**Introduction**

The interaction of light and matter is actively researched in last decades [1, 2]. The strength of the interaction is described by Rabi frequency. Placing particles into a cavity leads to enhancement of this interaction, and, when Rabi frequency becomes greater than damping rates in the system, transition to the strong coupling regime occurs. The latter leads to Rabi splitting [3, 4]. Beginning from realization in microwave range [5, 6], the strong coupling regime was later observed at optical frequencies in various systems, in particular in Fabry-Perot cavities [7-9], photonic crystal [10, 11] and semiconductor [12] cavities, plasmonic nanoresonator [13] strongly interacting with J-aggregated molecules [14-16], dye [17, 18] and photochromic [19] molecules.

Strong coupling leads to the formation of new energy states in the systems, resulting in change in some properties, including chemical reaction rates [20, 21], energy transfer rates [1, 22], and, last but not least, emission characteristics [23], including asymmetric spectrum shape [9, 22, 24, 25]. For this reason strong coupling attracts a lot of attention [1, 26] finding quantum [27-29], chemical and biological applications [30-32], enhanced optical processes, for example, enhanced Raman scattering and enhancement of the nonlinear refractive index [33-36]. Notably, interaction of the electric field of a plasmon resonator with molecules creates two hybrid states named exciton-polaritons [1, 2, 22, 37-39] (also known as exciton-polaritons and plexcitons). Polaritons can display fascinating behavior, such as non-equilibrium Bose-Einstein condensation [40, 41], and affect photochemical properties, like selective quenching or improvement of the quantum yields [23, 42]. However, in all those cases, strongly coupled systems are subject to a low intensity pump rate, so that only several of the lowest energy levels are excited [9, 39, 43, 44].

Subjecting matter to periodic external drive is an additional tool to control quantum systems. Such a control is referred to as Floquet engineering [45]. Renewed interest in Floquet engineering encompasses a vast array of open driven systems, such as strongly correlated electron systems [46-49], electron-phonon systems [50, 51], cold atoms [52, 53]. In turn, Floquet engineering of strongly coupled systems has not been investigated yet. Thus, understanding how strongly coupled systems can be controlled by an intense driving coherent field can open many avenues for applications of those systems.

In this letter, we investigate the synergetic effect of light-matter strong coupling and Floquet engineering on quantum radiation properties by considering a system of a strongly coupled molecule and a single mode resonator subjected to high-intensity coherent pump. The latter is supposed to be strong enough to alter the energy structure of the system. We develop a theory that enables the description of a high-intensity coherent pump and study the properties of the system, in particular, emission intensity and spectra. We show that high pump intensity can lead to substantially non-monotone dependence of luminescence intensity, excitation numbers and spectral width on pump field intensity. Moreover, we observe resonance-like peaks in luminescence intensity dependence on pump intensity. We show that the effect is caused by hybridization of initially orthogonal energy states in the system and can be tuned by changing the properties of external drive (pumping frequency and intensity). Our findings open new ways to experimentally control the internal properties of the system in the strong coupling regime by tuning pumping field frequency and amplitude. This in turn paves the way for additional control and tailoring of optical systems that already exploit or intend to exploit strong coupling, such as the creation of quantum devices, optical sensors, etc.

**Methods**

In our work, we consider high-intensity optical pumping of a strongly coupled system. We use the global approach based on the master equation for the density matrix in the Lindblad form [54, 55] to describe the properties of this system. We study the system of cavity coupled to molecule under coherent laser pump. In the rotating wave approximation [56, 57], this system is described by Hamiltonian

$$\hat{H} = \hbar\omega_0 \hat{a}^\dagger \hat{a} + \hbar\omega_0 \hat{\sigma}^\dagger \hat{\sigma} + \hbar\Omega_R(\hat{a}^\dagger \hat{\sigma} + \hat{a}\hat{\sigma}^\dagger) + \hbar\nu(\hat{a}^\dagger \exp(-i\omega t) + \hat{a}\exp(i\omega t)) \qquad (1)$$

Here $\hat{a}^\dagger$ and $\hat{a}$ are the creation and annihilation operators for photons in cavity mode. For the sake of simplicity, the molecule is described as a two-level system with $\hat{\sigma}^\dagger$ and $\hat{\sigma}$ being the raising and lowering operators between two levels. $\omega_0$ is the frequency of cavity mode and transition frequency of the molecule that are considered to be equal. $\Omega_R$ is the coupling constant between cavity mode and molecule (Rabi frequency [56]). $\nu$ is the coupling constant between the pump light and the molecule, proportional to the amplitude of the incident field. $\omega$ is the pump frequency.

The first three terms of Eq. (1) are the Jaynes-Cummings Hamiltonian with equal transition frequencies for molecule and cavity [56, 58]. The eigenstates of the Jaynes-Cummings Hamiltonian are well-known and include the ground state where both the cavity mode and molecule are unexcited, and an infinite set of polariton pairs with eigenstates $(|n,g\rangle \pm |n-1,e\rangle)/\sqrt{2}, n=1,2,...$ [58, 59]. As evident from the eigenstates, polariton is a hybrid state between photons of the cavity and the excited molecule (exciton). The energy of polaritons is $n\omega_0 - \sqrt{n}\Omega_R$ for $n$-th lower polariton (LP$n$) and $n\omega + \sqrt{n}\Omega_R$ for $n$-th upper polariton (UP$n$) [58]. Finally, the last term in Eq. (1) is the interaction of the pump field with the cavity mode. Here the interaction of the pump field with the cavity is assumed to be stronger than the interaction of pump field with the molecule's dipole; therefore, the latter is neglected.

Description of quantum of system dynamics requires two more steps: 1) accounting for relaxation processes through respective reservoirs, 2) eliminating time dependence from the Eq. (1). These details are discussed in Supplementary Materials A. The final equation that describes the behavior of the system then becomes

$$\frac{\partial \hat{\tilde{\rho}}}{\partial t} = \frac{i}{\hbar}\left[\hat{\tilde{\rho}}, \hat{H}_{system}\right] + \hat{L}_a(\hat{\tilde{\rho}}) + \hat{L}_\sigma(\hat{\tilde{\rho}}) + \hat{L}_{deph}(\hat{\tilde{\rho}}) \qquad (2)$$

$$\frac{\hat{H}_{system}}{\hbar} = \Delta \hat{a}^\dagger \hat{a} + \Delta \hat{\sigma}^\dagger \hat{\sigma} + \Omega_R(\hat{a}^\dagger \hat{\sigma} + \hat{a}\hat{\sigma}^\dagger) + \nu(\hat{a}^\dagger + \hat{a}) \qquad (3)$$

$\Delta = \omega_0 - \omega$ is the difference between the cavity mode frequency and the external drive frequency. $\hat{\tilde{\rho}}(t)$ is the density matrix of the system. It describes probabilities and correlations between the states described by the Hamiltonian (3). $\hat{L}_{deph}(\hat{\rho})$ describes dephasing process. $\hat{L}_a(\hat{\rho})$ describes radiative and non-radiative relaxations of the cavity mode. $\hat{L}_\sigma(\hat{\rho})$ describes radiative and non-radiative longitudinal relaxations of molecule. Crucially, we use the global approach to describe the relaxation [60]. This means that operators $\hat{L}_{deph}(\hat{\rho})$, $\hat{L}_a(\hat{\rho})$, $\hat{L}_\sigma(\hat{\rho})$ consist of terms proportional to transitions between eigenstates of $\hat{H}_{system}$. This is distinct from the local approach [54, 55, 60] where relaxation processes are described through terms proportional to eigenstates of operators $\hat{a}^\dagger \hat{a}$ and $\hat{\sigma}^\dagger \hat{\sigma}$. The global approach is necessary in our case since the Rabi constant $\Omega_R$ is greater than relaxation rates in the strong coupling regime that we aim to describe. A detailed examination of different approaches to relaxation description and their applicability can be found in [54, 55, 60].

We use the Eq. (2) to find the stationary density matrix. Average values of observable parameters $f$ can be found using an expression $\langle \hat{f} \rangle = Tr(\hat{\rho}_{st}\hat{f})$.

### Results

We start by examining energy levels of the effective Hamiltonian (3) at different incident field amplitudes, $\nu$. Relaxation rates used in the consideration are $\gamma_a = 10^{-3}\omega_0$, $\gamma_D = 10^{-5}\omega_0$,

$\gamma_{ph} = 5 \times 10^{-3} \omega_0$. The temperature $T = 0.02\omega_0$ is taken into account in the dephasing rate, $\gamma_{ph}$, and its value is about the room temperature. The coupling constant between the molecule and the cavity mode is $\Omega_R = 0.01\omega_0$. The pumping frequency is $\omega = 1.0101\omega_0$, which is slightly above the transition frequency between the ground state and UP in a system without external drive, $\omega_{UP1} = \omega_0 + \Omega = 1.01\omega_0$. This is one of the common ways to realize coherent pump experimentally to avoid mixing radiated and scattered light from the system [61-63]. Detailed parameters of the cavity mode and the active molecule that could facilitate those rates are discussed in Supplementary Materials B.

Numerically solving the Eq. (2) we find the stationary density matrix of the system. From the density matrix, we can obtain all the information about the stationary states of the system. In particular, we find a dependence of the number of excitations in cavity mode on the amplitude pump rate using expression $n_{st} = Tr(\hat{a}^\dagger \hat{a} \hat{\rho}_{st})$ [Figure 1(a)]. Our calculations show that there are several peaks at pump rates $\nu = 2.8 \times 10^{-3} \omega_0$, $\nu = 4.9 \times 10^{-3} \omega_0$, $\nu = 9.2 \times 10^{-3} \omega_0$, etc. We also find energy flow related to radiative relaxation of the cavity mode [Figure 1(b)] (total energy flow to the reservoir corresponding to the free space, $Tr(\hat{L}_a(\hat{\rho}_{st}))$). It displays peaks near similar pump rate values.

Thus, unexpectedly, increasing the pumping field amplitude can lead to a decrease in luminescence and excitation numbers. This is in stark contrast to monotone behavior observed at low pump rates [64, 65]. Additionally, the emission spectra become wider in the vicinity of peaks; see Figure 2, which illustrates linewidth near pump rate $\nu = 4.9 \times 10^{-3} \omega_0$.

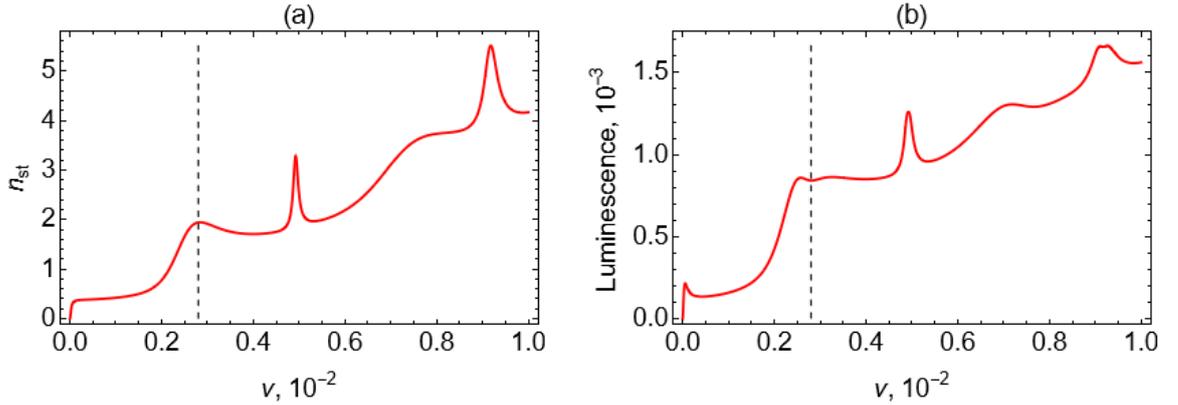

Figure 1. (a) Dependence of the number of excitations, $n_{st}$, in the cavity mode on the amplitude of pump rate, $\nu$, (b) The dependence of the energy flow related to radiative and non-radiative relaxations of the cavity mode on $\nu$. The dashed line corresponds to $\nu = 2.8 \times 10^{-3}$.

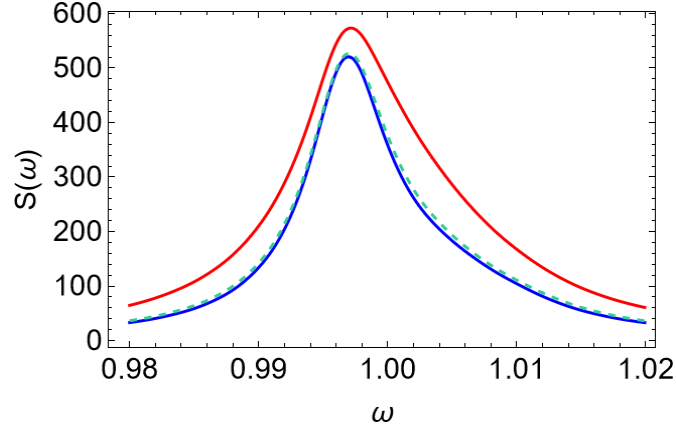

Figure 2. Emission spectrum of cavity mode (the blue line is calculated for $v = 4.5 \times 10^{-3} \omega_0$, the red line is calculated for $v = 4.9 \times 10^{-3} \omega_0$, the green is calculated for $v = 5.3 \times 10^{-3} \omega_0$). The emission line width is the largest near $v = 4.8 \times 10^{-3} \omega_0$, at which the dependence of the number of excitations on the pump rate has the local maximum [Figure 1b].

To explain the non-monotone luminescence and linewidth behavior in the system with strong pumping, we are going to investigate relaxation processes near the first peak at $v = 2.8 \times 10^{-3} \omega_0$. To understand the behavior of the system, we examine eigenstates of the Hamiltonian Eq. (3). We can see that near $v = 2.8 \times 10^{-3} \omega_0$ a couple of energies are located extremely close to each other (blue - LP with one excitation, which we denote as LP1, red - UP with four excitations, which we denote as UP4), see Figure 3. These levels are the closest at the same pump rate as the excitation number maximum is observed (see Figure 4). We also find that near this pump rate, LP1 and UP 4 states hybridize, and this hybridization allows the population of polaritons with high excitation numbers, which then undergo cascade radiative transitions into the ground state (more details in Supplementary Materials C), giving rise to the non-monotone behavior of luminescence, spectral width and excitation numbers associated with state hybridization. Similarly, other peaks in Figure 1 are also associated with hybridization of other pairs of levels.

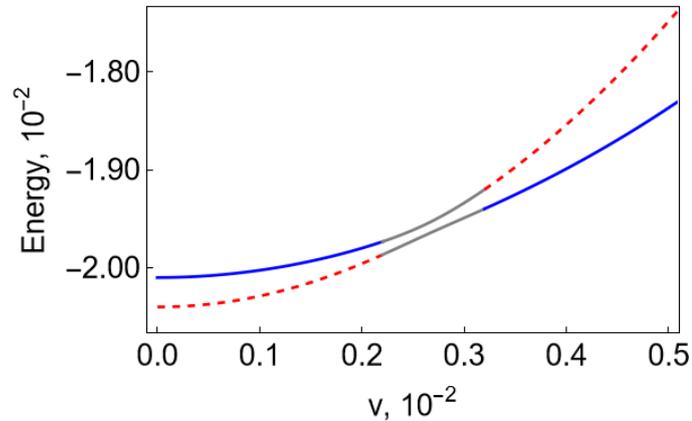

Figure 3. Dependence of eigenenergies of the effective Hamiltonian on the coupling constant between the pump light and the molecule $v$. The energies corresponding to LP1 (blue curve) and UP4 (red curve).

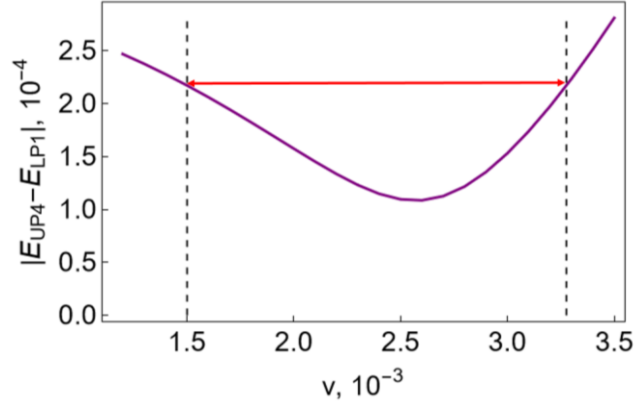

Figure 4. Dependence of differences between energies of levels UP1 and LP4 on $\nu$. The width of the peak, measured at twice the lowest energy difference, is $\Delta \nu = 1.8 \times 10^{-3} \omega_0$.

With all the observations in mind, we deduce the mechanism for the non-monotone behavior of system characteristics. The simplified scheme of energy levels for peaks near $\nu = 2.8 \times 10^{-3}$ is shown in Figure 5 (for changes in transition rates, see Figure 1S in Supplementary Materials C). Note that the original Hamiltonian $\hat{H}$ is time-dependent, and thus we instead show the energy levels of the transformed time-independent Hamiltonian $\hat{H}_{system}$. The system exhibit hybridization of levels near the pump rate $\nu = 2.8 \times 10^{-3}$, but it becomes less pronounced the further away we change the pump rate, $\nu$. Without level hybridization, the coherent pump rate aided by dephasing leads to transitions from ground state to UP1 state (without dephasing the transition behaves as coherent drive rather than pumping [66]). Then the population of UP1 level can radiatively degrade into the ground state or non-radiatively transit into LP1 due to dephasing (this is a well known mechanism in systems with strong coupling, resulting in upper polariton being dark [9, 39, 67]), which in turn radiatively degrades into the ground states. See the respective sketch in Figure 5. This is the system's "normal" behavior, typical for the low pump rate limit. However, the structure changes near the pump rate $\nu = 2.8 \times 10^{-3}$. Because two levels, LP1 and UP4, mix, many more additional transitions become possible. Population can now transit from UP1 to a pair of levels that are a mixture of LP1 and UP4, owing to the transition $UP1 \to LP1$. Therefore, in the system with hybridized levels a new cascade of radiative transitions between upper polaritons arises (UP4 to UP3 to UP2 to UP1 to ground state; the last one was already present). Further increasing the effect is the fact that the mixture of UP4 and LP1 can transit to LP4 due to dephasing process $UP4 \to LP4$, which produces another cascade of transitions (LP4 to LP3 to LP2 to LP1 to ground state; the last one was already present). Additional transitions, such as UP3 to LP3, also occur, but they are omitted from the illustration for the sake of clarity. We call this state mixing Floquet hybridization, as it requires an external periodic drive to arise. Note that this is distinct from molecule-field hybridization in polaritons in the Jaynes-Cummings model. Similar Floquet hybridization and the resulting cascade of radiative transitions occur near other pump rates $\nu = 4.9 \times 10^{-3} \omega_0$, $\nu = 9.2 \times 10^{-3} \omega_0$ where other luminescence peaks are observed.

In other words, Floquet hybridization of an upper polariton and a lower polariton with different excitation numbers facilitates direct pumping of levels with large numbers of excitations, from levels with low number of excitations. In contrast, without Floquet hybridization the only way to pump high excitation number polariton is to subsequently pump all polaritons. Direct pumping of polaritons with high excitation number occurs at the cost of the population of polaritons with low excitation numbers, thus increasing overall excitation numbers in the system. Since levels with higher excitation numbers are pumped, the luminescence of the systems also grows (explaining Figure 1). And since more different radiative transitions with various frequencies occur, the spectral width increases when the levels are hybridized (explaining Figure 2). Note that in this discussion we have intentionally avoided non-radiative relaxation of the cavity mode and the molecule. This is because these transitions always accompany radiative transitions and do not qualitatively affect the behavior of energy levels and population dynamics. Their values are relevant to luminescence characteristics (such as ones shown in [Figure 1 and 2]), but not to the essence of the effect.

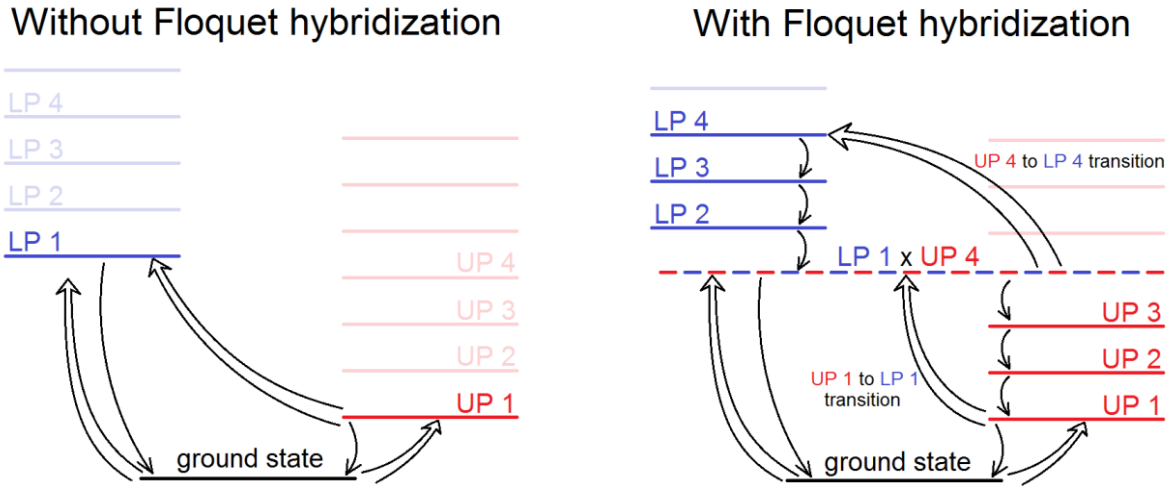

Figure 5. Simplified scheme of transition between energy levels of $\hat{H}_{system}$ (see Eq. (3)), caused by dephasing (double arrows) and radiation transition (single arrows). Slow transition rates, and transitions from unpopulated levels are omitted for the sake of clarity. Levels with low population are greyed out. (a) The transitions in the system without Floquet hybridization between upper and lower polaritons. Only the ground state, UP 1 and LP 1 are significantly populated, so other levels have negligible effect on the behavior of the system. (b) The transitions in the system with Floquet hybridization between upper and lower polaritons. There are significantly more radiative transitions (single arrows) in the system with hybridization, because more energy levels are populated. Note that the levels depicted here are levels of the effective Hamiltonian, $\hat{H}_{system}$ (Eq. 3), not of the original Hamiltonian $\hat{H}$ (Eq. 1).

Let us discuss the conditions under which the effect arises. The position of peaks in Figure 1(a-b) is determined by the combination of parameters for which the appropriate pair of LP and UP with non-zero population reaches similar energies and hybridizes. Since level structure is determined by the Hamiltonian, $\hat{H}_{system}$ (Eq. (3)), the position of peaks depends on three parameters: the coupling constant between the cavity mode and molecule, $\Omega_R$; the coupling constant between the pump light and the molecule, $\nu$, and the pump frequency detuning between the cavity mode frequency and external drive frequency, $\Delta$, and does not depend on relaxation rates. Whereas the exact criterion for appearance of the effect remains elusive, we find that since many levels in a hybrid system can be excited at the same time, it is easy to numerically find parameters where the effect is noticeable. To illustrate this, we plot the excitation number in the parameter space of coupling to external drive, $\nu$, and pump frequency, $\omega$, [see Figure 6]. At least three near-coalescences can be seen (curves where the excitation number is higher). A similar picture appears for other values of pumping frequency, $\omega$, provided it lies between transition frequencies of unperturbed LP 1 ($\omega_0 - \Omega_R$) and UP 1 ($\omega_0 + \Omega_R$).

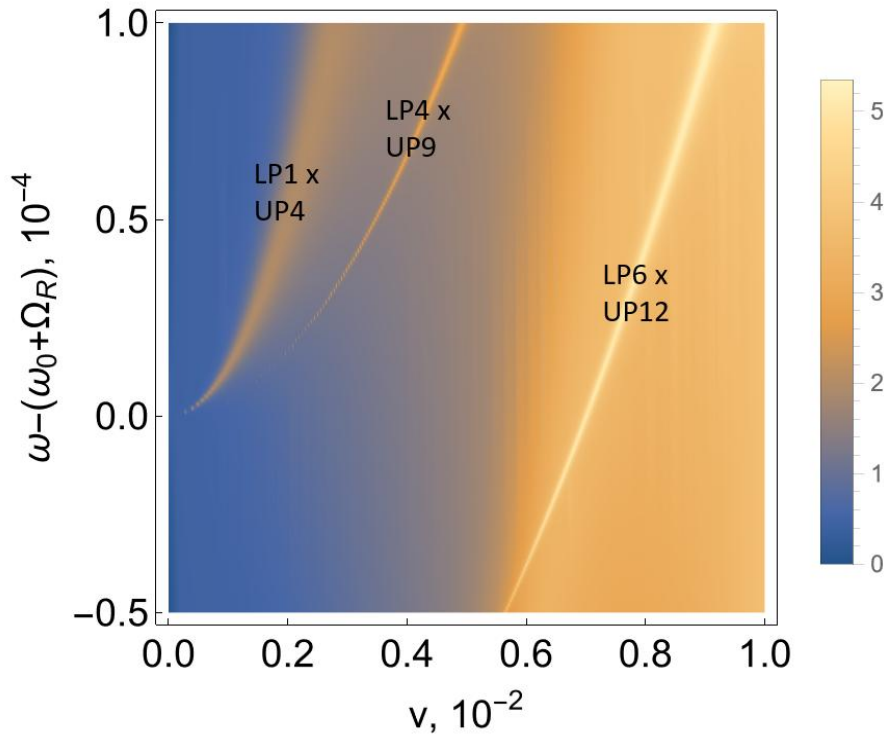

Figure 6. Number of excitations plotted against pump rate $v$ (horizontal axis) and pump frequency detuning from the first upper polariton state $\omega - (\omega_0 + \Omega_R)$ (vertical axis). The ridges correspond to hybridization of different pairs of levels.

**Discussion and Conclusion**

We have developed a theory that describes a hybrid system of a molecule coupled to a cavity mode in the strong coupling regime subjected to high-intensity pump rate. In our consideration, the pump rate is comparable to the coupling constant, which results in a substantial change of energy structure of the system. We show that a high pump rate results in the mixing of energy states in the system near certain combinations of pump rate and pumping frequency. We call this mixing Floquet hybridization, as the periodic drive is necessary to achieve it. Floquet hybridization facilitates direct pumping from the ground state to energy levels with a high number of excitations, allowing the system to absorb more energy from the pump field. As a result, more radiative transitions occur in the system, giving rise to a sharp increase in luminescence intensity and an increase in the spectral width of the emitted light. Notably, the effect is substantially non-monotone, i.e., an increase in pump rate drives the system out of the optimal parameters for hybridization, and the luminescence intensity drops. Thus, we demonstrate a new way to employ Floquet engineering to alter the energy structure of a hybrid system.

These results offer new tools to engineer strong coupling systems and to control their properties via external drive. With simple modification, this theory can be applied to a system containing many molecules. The sharp change in luminescent characteristics occurring near specific values of easily controllable parameters (optical pump intensity and optical pump frequency) can be of great use to designing non-linear optical elements. We believe that the ability to engineer the properties of the system using an external pump through energy level manipulation is not limited to hybrid cavity-molecule systems and can be applied to other coupled optical systems. Strong non-linear optical effects emerging from the proposed configuration can be of great aid to the development of optical computers [38, 68, 69] and other optical devices [61-63].

## Acknowledgment

The study was financially supported by a Grant from Russian Science Foundation (Project No. 23-42-00049). I.V.D. and E.S.A. thank Advancement of Theoretical Physics and Mathematics Foundation «Basis».